\documentclass[a4paper,11pt]{article}
\usepackage{jinstpub} 
\usepackage{lineno}
\usepackage{booktabs}
\usepackage{multirow}
\usepackage{subcaption}
\usepackage{siunitx}

\proceeding{21$^{\text{st}}$ Workshop on Advanced Silicon Radiation Detectors -- TREDI\\
2026-02-17 to 2026-02-19\\
Perugia, Italy}

\title{\boldmath Development of Segmented 4H-SiC LGADs}

\author[a]{Vojtěch Kráčmar,} 
\author[b]{Jan Chochol,}
\author[b]{Adam Klimsza,}
\author[c]{Jana Kozáková,}
\author[b]{Adam Kozelsky,}
\author[c]{Jiří Kroll,}
\author[d]{Adela Kubránska,}
\author[a]{Mária Marčišovská,}
\author[c]{Marcela Mikeštíková,}
\author[a]{Radek Novotný,}
\author[e]{Aymeric Privat,}
\author[b]{Peter Slovák,}
\author[a]{Tobiáš Vasiljev,}
\author[a,c,1]{Peter Švihra\note{Corresponding author.}}

\affiliation[a]{Department of Physics, FNSPE CTU in Prague, Brehova 78/7, Prague, 115 19, Czechia}
\affiliation[b]{onsemi, 1. máje 2230, Rožnov pod Radhoštěm, 756 61, Czechia}
\affiliation[c]{Department of Detector Development and Data Processing, FZU CAS, Na Slovance 1999/2, Prague, 182 00, Czechia}
\affiliation[d]{onsemi, Twin City C, Mlynské nivy 14, Bratislava, Slovakia}
\affiliation[e]{onsemi, 5701 N Pima Road, Scottsdale AZ 85250, USA}

\emailAdd{peter.svihra@cern.ch}

\abstract{The wide-bandgap semiconductor 4H-silicon carbide (4H-SiC) offers a compelling combination of radiation hardness, thermal stability, and high critical electric field for particle detection in harsh environments.
To compensate for the comparatively low charge generation in SiC, the Low-Gain Avalanche Detector (LGAD) concept can be adopted to provide internal signal amplification.
Building on three preceding generations of single-pad 4H-SiC LGAD prototypes fabricated by ion implantation, this work presents the design, fabrication, and initial characterization of segmented 4H-SiC LGAD devices -- the first fabricated and characterized devices reported.
Strip detectors with \SI{80}{\micro\meter} pitch and pixel arrays with 55 and \SI{110}{\micro\meter} pitch were produced using multiple inter-channel isolation strategies, including geometric separation and oxide-filled trenches.
Two-photon absorption transient current technique (TPA-TCT) measurements performed at ELI ERIC demonstrate clear charge separation between adjacent strips with internal gain, confirming functional segmentation.}


\begin{document}
\maketitle
\flushbottom

\section{Introduction}
\label{sec:introduction}

The increasing demand for radiation-hard, fast-timing detectors in collider experiments, space instrumentation, and nuclear applications has driven a sustained search for sensor materials capable of reliable operation in environments where conventional silicon devices struggle. Within this landscape, wide-bandgap semiconductors have emerged as a particularly promising class, and 4H-silicon carbide (4H-SiC) -- the most technologically mature of the SiC polytypes -- has become a focus of recent detector R\&D. Its appeal has been further reinforced by improvements in industrial wafer quality and steadily decreasing costs, both consequences of the rapid expansion of the power electronics market.

The intrinsic properties of 4H-SiC are well suited to this role~\cite{levinshtein2001properties}. Its bandgap of approximately 3.26 eV -- close to three times that of silicon (1.12 eV) -- together with a critical electric field roughly an order of magnitude higher than Silicon, supports operation at high bias voltages and at elevated temperatures. This also brings the added benefit of negligible sensitivity to visible light. Carrier saturation velocities are large for both electrons ($\mathrm{\approx 1.47\times 10^7 cm/s}$) and holes ($\mathrm{\approx0.69 \times 10^7 cm/s}$)~\cite{gsponer2025extraction}, and displacement threshold energies are correspondingly high, with values around 21 eV for carbon and 35 eV for silicon in the SiC lattice~\cite{rafi2020electron, moscatelli2026radiation}, underpinning the material's expected resistance to bulk damage. A further consequence of the wide bandgap is an intrinsic carrier concentration that lies roughly eighteen to nineteen orders of magnitude below that of silicon, so that thermally generated leakage currents remain modest even after being exposed to substantial radiation doses. The thermal conductivity of 4H-SiC, approximately 370~W/(m$\cdot$K) at room temperature~\cite{wei2013thermal}, exceeds that of silicon by more than a factor of two, facilitating heat dissipation in high-rate environments.

These advantages, however, are accompanied by an inherent limitation. Because of its larger bandgap, 4H-SiC produces only about of 57 electron–hole pairs per micrometer for a minimum ionizing particle~\cite{christanell2022hsilicon}, and currently available epitaxial layers rarely exceed 100 $\mathrm{\upmu m}$ in thickness. The combination yields signals that are comparatively small for precision timing or tracking applications. A natural way to compensate is to introduce internal charge multiplication, following the well-established Low-Gain Avalanche Detector (LGAD) concept developed in silicon\cite{pellegrini2014technology}, in which a highly doped gain layer enables controlled avalanche amplification of the primary signal. Two distinct routes have been explored to obtain such a gain region in 4H-SiC: epitaxial growth of a buried doped layer and ion-implantation-based definition of the multiplication profile, the latter mirroring the established workflow used in silicon detector fabrication.

In silicon, the LGAD concept has matured rapidly since its introduction~\cite{pellegrini2014technology}, driven in particular by its adoption for the ATLAS High-Granularity Timing Detector (HGTD)~\cite{atlas2020hgtd} and the CMS Endcap Timing Layer (ETL). Segmenting the gain layer into individual pads, however, introduced a well-known challenge: the gain implant must be terminated at each pad boundary, creating an inactive ``no-gain region'' that reduces the effective fill factor~\cite{paternoster2021novel}. Several mitigation strategies have been developed in silicon, including optimized junction termination extensions (JTE), p-stop implants, oxide-filled trench isolation (TI-LGAD), and resistive AC-coupled readout (AC-LGAD/RSD)~\cite{paternoster2021novel}. The segmented 4H-SiC LGAD devices presented in this work face the same fundamental challenge, and the isolation strategies explored here -- geometric separation and trench isolation -- are informed by this silicon experience.

First-generation 4H-SiC LGAD prototypes have already demonstrated measurable internal gain and encouraging electrical and signal-formation behavior, as documented in~\cite{novotn2025first, svihra2025exploring, zhao2024electrical, yang2026characterization}. Independently, segmented 4H-SiC PN detectors without internal gain have been reported~\cite{zatko2022from}. The natural next step is to combine both achievements: internal charge multiplication and multi-channel segmentation in a single device. No such segmented 4H-SiC LGAD has been demonstrated to date. This work presents the first: segmented 4H-SiC LGADs fabricated by ion implantation, covering both strip and pixel geometries, together with their initial characterization.

\section{Overview of 4H-SiC detector developments}
\label{sec:current_status}

This section summarizes the principal 4H-SiC detector developments that inform the present work: groups pursuing internal signal amplification in single-pad LGADs, and the demonstration of segmented 4H-SiC PN pixel detectors without a gain layer. The 4H-SiC devices presented in this work combine both achievements -- internal gain and segmented geometry.

\paragraph{SICAR} A 4H-SiC LGAD device (named SICAR for SIlicon CARbide) was produced and tested by Zhao et al.~\cite{zhao2024electrical} using the epitaxial growth approach. A \SI{50}{\micro\meter} thick n-type epitaxial drift layer is grown on an n$^{++}$ substrate with an n buffer layer in between. A \SI{1}{\micro\meter} thick nitrogen-doped n$^{+}$ gain layer and a \SI{0.3}{\micro\meter} aluminium-doped p$^{++}$ anode layer are grown on top, and the stack is shaped into circular mesa structures (diameter \SI{1000}{\micro\meter} ) by selective etching down to the n-epi. A SiO$_2$ passivation layer covers the bevel edge. Tests with an $^{241}$Am source demonstrated a gain factor of approximately~3 at 300~V reverse bias relative to a reference PN device fabricated without the gain layer.

\paragraph{LBNL} Yang et al.\ used a similar mesa-based design~\cite{yang2026characterization}. Their device features a thinner gain layer (\SI{0.5}{\micro\meter}) and a thicker drift region (\SI{75}{\micro\meter}), with circular mesa diameters ranging from \SIrange{75}{600}{\micro\meter}. Charge multiplication was confirmed using $\upalpha$ particles from a $^{210}$Po source, yielding a gain of approximately 2-3. In subsequent work~\cite{yang2025ultrafast}, metallic field plates at the bevel edge were introduced to increase the breakdown voltage. UV TCT measurements with a 1~MIP-equivalent laser signal yielded a gain factor of 7-8 at 500~V bias.

\paragraph{STU} Zaťko et al.\ investigated epitaxially grown 4H-SiC PN diodes~\cite{zatko2022from}, demonstrating good spectral resolution and charge collection efficiency with combined $\upalpha$-particle sources ($^{238}$Pu, $^{239}$Pu, $^{244}$Cm). Notably, the authors also presented a $256\times256$ pixel matrix with \SI{80}{\micro\meter} epitaxy and \SI{55}{\micro\meter} pitch bump-bonded to a Timepix3-based readout, representing an early segmented 4H-SiC radiation detector -- albeit without a gain layer.

\paragraph{RD50-SiC-LGAD common project} The CERN-driven RD50 and subsequent DRD3 WG6 collaborations contributed a detailed design and simulation study published by Onder et al.~\cite{onder2025design}. Their approach combines epitaxial growth with ion implantation: a \SI{27.6}{\micro\meter} n-epi layer is grown on the substrate, separated by a \SI{1}{\micro\meter} n$^{++}$ buffer, and a \SI{2.4}{\micro\meter} gain layer is grown covering the full wafer area. The p$^{++}$ anode is then implanted. Inter-device isolation is achieved by a combination of \SI{7}{\micro\meter} deep etched trenches and implanted JTE. No fabricated devices have been reported to date; simulations predict a gain factor of 1-10 depending on bias voltage, full depletion around 500~V, and breakdown above 2.4~kV.

\paragraph{CAPADS} The collaboration of the Center of Applied Physics and Advanced Detection Systems (CAPADS) at the Faculty of Nuclear Sciences and Physical Engineering, Czech Technical University in Prague (FNSPE CTU), the Institute of Physics of the Czech Academy of Sciences (FZU) and the onsemi company pursues an ion-implantation-based approach optimized for onsemi's fabrication facility in Rožnov pod Radhoštěm, Czech Republic. The LGAD devices are built on commercially available 6-inch 4H-SiC n-type wafers (substrate doping $\sim 10^{19}$~cm$^{-3}$) with epitaxial layers of \SIrange{30}{100}{\micro\meter} thickness and low doping concentration (down to approximately $5\times10^{13}$~cm$^{-3}$). Individual diodes occupy an area of roughly $3\times3$~mm$^2$, electrically terminated by aluminium-implanted JTEs designed for breakdown voltages above 1~kV. The gain layer is formed by nitrogen implantation, with the resulting multiplication factor controlled through variation of implant energies and doses. An exhaustive description of the design is provided in~\cite{novotn2025first}.

Three production batches (Lots~1-3) of single-pad devices were fabricated between 2022 and 2025, covering epitaxial layer thicknesses of \SIlist{30;50;100}{\micro\meter} and two LGAD implant variants per Lot targeting different gain factors. Electrical characterization (IV, CV), UV TCT, $\upbeta$-source measurements ($^{90}$Sr), and irradiation studies with protons, neutrons, and gammas were performed across these Lots. The results -- including production yields of approximately 85\%, gain factors around 20, and promising sub-100~ps timing resolution -- are detailed in~\cite{novotn2025first, svihra2025exploring}, with radiation hardness results forthcoming. The fourth batch (Lot~4), whose production began in 2025, introduces the first segmented device geometries and forms the subject of this work.

\begin{figure} [t]
    \centering
    \includegraphics[width=1\textwidth]{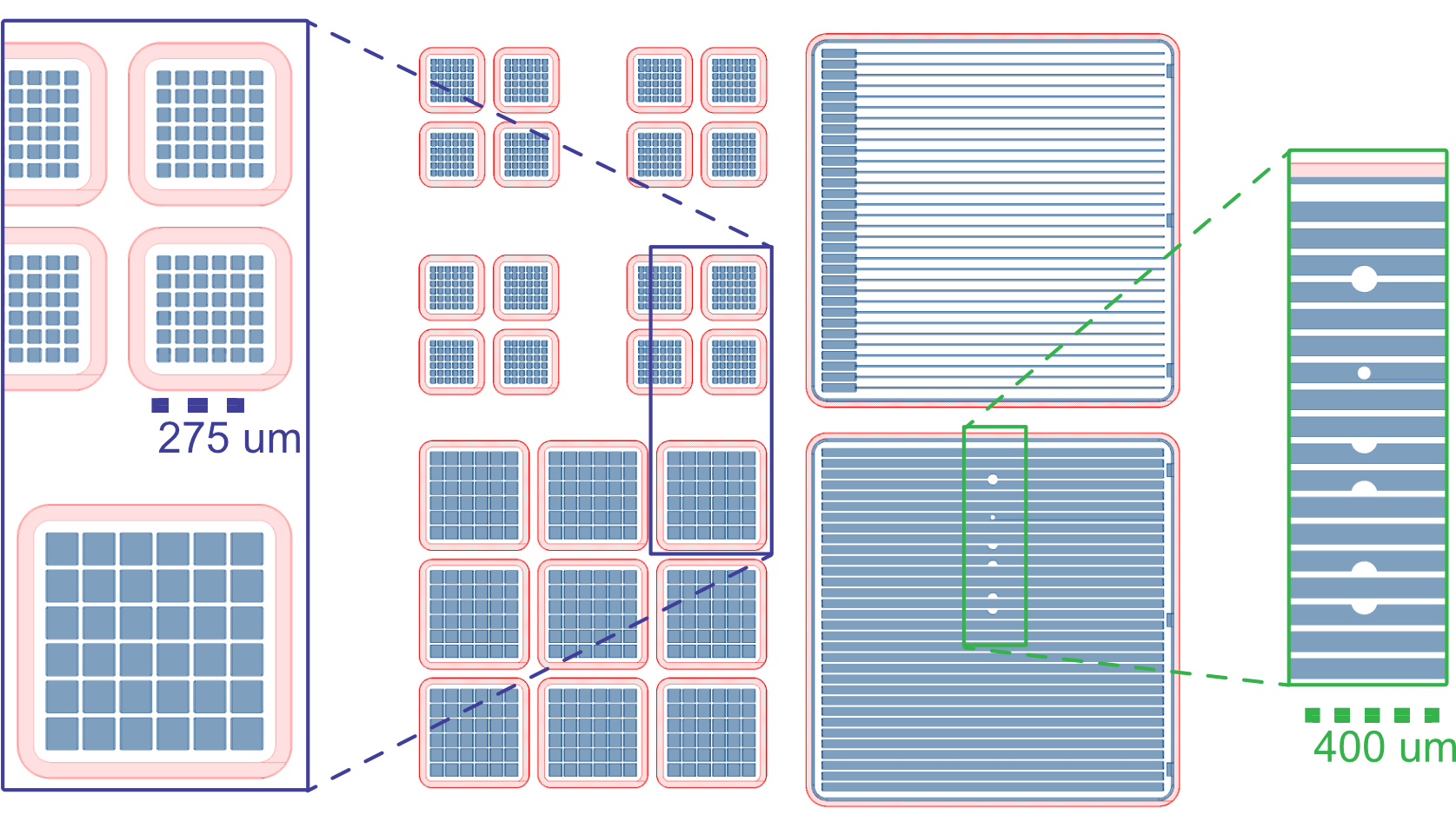}
    \caption{Layout drawings of four of the device variants investigated in
this work, extracted from the wafer shot map. The two dies on the
right are strip sensors differing in the width of the top metal
cover, while the two dies on the left are pixel sensors in
the \emph{Split} configuration introduced in Section~\ref{sec:segmented_design}.
The top metal is shown in blue and the JTE in red. The blue inset (left) shows a magnified
view of a Split pixel cell; the green inset (right) shows a section
of the strip array around the openings in the top metal
used for TCT laser access.}
    \label{fig:die_schematic}
\end{figure}

\section{Segmented LGAD design}
\label{sec:segmented_design}
The control over the implantation process and the fabrication yield demonstrated during Lots~1--3 provided a firm basis to proceed to segmented 4H-SiC LGAD prototypes. Lot~4 marks the first major departure from the single-pad design: the variety of manufactured devices was expanded to include strip and pixel geometries, targeting compatibility with established readout systems -- specifically ALIBAVA~\cite{marco2009alibava} for strips and Timepix4~\cite{llopart2022timepix4} for pixels.

\subsection{General design split}
The fabrication layout accommodates 49 individual dies per shot, with each die including markings indicating the die number and principal design parameters, allowing reliable post-dicing identification under microscopic inspection.
Of these, 45 are dedicated to different device designs and the remaining 4 to production monitoring structures. 
The split covers both LGAD and reference PN structures to enable direct comparison.
Two LGAD implant variants were used, consistent with the previous Lots. Each device occupies approximately $3\times3$~mm$^2$.

\paragraph{Strips} Two metal contact variants of strip detectors were designed, both consisting of 32 rectangular strips covering the full active length of the sensor with \SI{80}{\micro\meter} pitch. The first variant (referred to as ``Strip \SI{60}{\micro\meter}'') features a metallization layer covering the strip fully in a width of \SI{60}{\micro\meter}, with circular and semicircular openings etched at multiple positions to allow laser light to reach the bare SiC surface for TCT studies (figure~\ref{fig:die_schematic}, bottom right). The second variant (``Strip \SI{5}{\micro\meter}'') uses a narrow \SI{5}{\micro\meter} metal contact running most of the strip length, combined with a $\qtyproduct{60 x 260}{\micro\meter}$ bond pad at one end (figure~\ref{fig:die_schematic}, top right).

\paragraph{Pixels} Pixel devices were designed in two pitches: \SIlist{55;110}{\micro\meter} for bump-bonding to Timepix4~\cite{llopart2022timepix4}, with the latter working also for manual wire-bonding (figure~\ref{fig:die_schematic}, bottom middle). These were further subdivided into full-die single-design variants and split dies combining multiple separation strategies to maximize the design coverage per shot.

\begin{table}[t]
  \centering
  \caption{Summary of the device geometries investigated in this work. Strip sensors share a common pitch of \SI{80}{\micro\meter}; the labels ``\SI{5}{\micro\meter}'' and ``\SI{60}{\micro\meter}'' refer to the width of the top metal cover. Pixel sensors are produced at two pitches \SIlist{55;110}{\micro\meter}. Strip sensors are fabricated as both LGAD and PN diodes, while pixel sensors are LGAD-only and are available in \emph{Full} and \emph{Split} pixel groupings. Two strategies are used to isolate neighboring electrodes: trench isolation and purely geometric separation. For each strategy the table reports the relevant spacing and gap parameters. Multiple comma-separated values in a single cell indicate that several variants of that geometry are present on the wafer; a dash ``--'' marks configurations that are not implemented.}
  \begin{tabular}{lllcccc}
    \toprule
     & \multirow{2}{*}{Grouping} & \multirow{2}{*}{Type} & \multicolumn{2}{c}{Trench} & \multicolumn{2}{c}{Separation} \\
    \cmidrule(lr){4-5} \cmidrule(lr){6-7}
     &  &  & Spacing [\si{\micro\meter}] & Gap [\si{\micro\meter}] & Spacing [\si{\micro\meter}] & Gap [\si{\micro\meter}] \\
    \midrule
    \multirow{2}{*}{Strip \SI{5}{\micro\meter}}  &       & LGAD & 0, 1       & 0, 1 & 0.75, 1.5 & 0, 1, 2.5 \\
                                        &       & PN   & 0, 1       & --   & 0.75, 1.5 & --        \\
    \midrule                                    
    \multirow{2}{*}{Strip \SI{60}{\micro\meter}} &       & LGAD & 0, 1       & 0, 1 & 0.75, 1.5 & 0, 1, 2.5 \\
                                        &       & PN   & 0, 1       & --   & 0.75, 1.5 & --        \\
    \midrule                                    
    \multirow{2}{*}{Pixel \SI{55}{\micro\meter}}  & Full  & LGAD & 1          & 1    & --        & --        \\
                                        & Split & LGAD & 0, 0.75, 1 & 0, 1 & 0.75, 1.5 & 0, 1, 2.5 \\
    \midrule                                    
    \multirow{2}{*}{Pixel \SI{110}{\micro\meter}} & Full  & LGAD & 1          & 1    & --        & --        \\
                                        & Split & LGAD & 0, 1       & 0, 1 & 0.75, 1.5 & 0, 1, 2.5 \\
    \bottomrule
  \end{tabular}
  \label{tab:lgad-split}
\end{table}

\begin{figure} [t]
    \centering
    \begin{subfigure}[t]{0.49\textwidth}
        \includegraphics[width=1.0\linewidth]{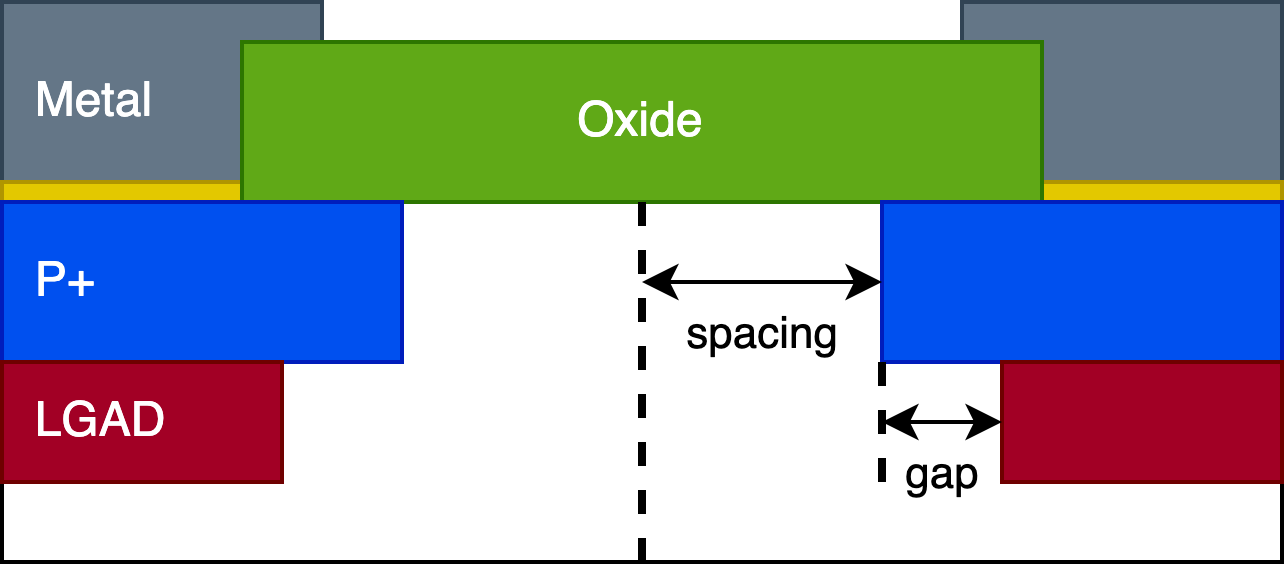}
        \caption{Spacing}
    \end{subfigure}
    \begin{subfigure}[t]{0.49\textwidth}
        \includegraphics[width=1.0\linewidth]{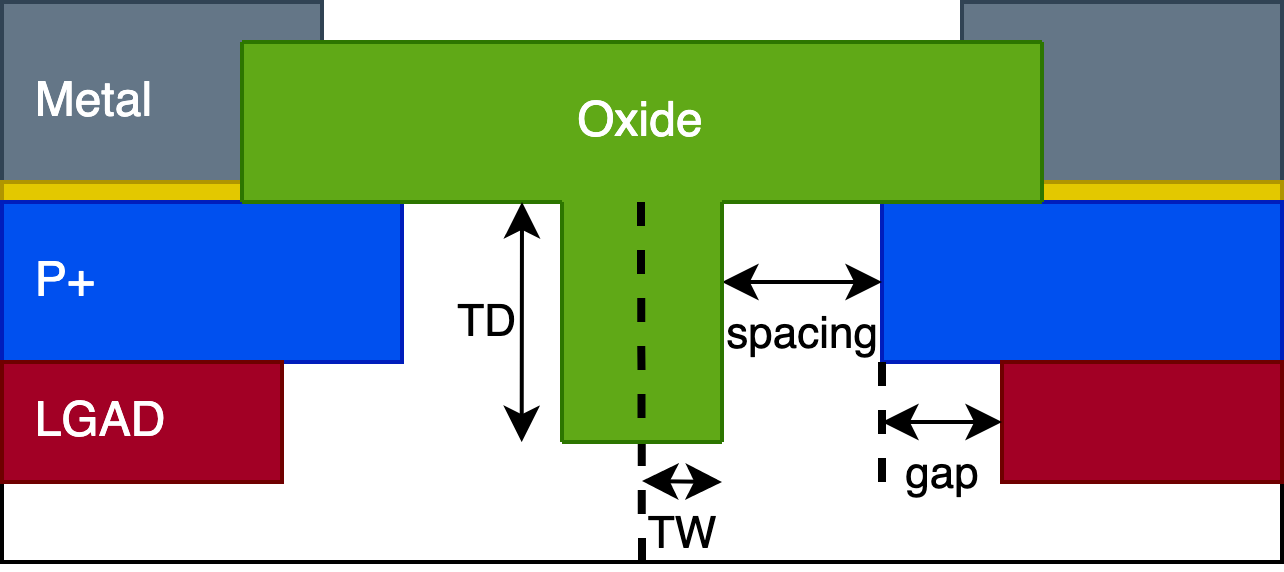}
        \caption{Trench}
    \end{subfigure}
    \caption{Schematic cross-sections of the two inter-channel isolation approaches used in the segmented 4H-SiC LGAD design. Left: geometric separation without trench -- the spacing parameter denotes the lateral distance from the p$^{+}$ implant edge to the geometric mid-plane (dashed line) between adjacent active regions, and the gap parameter denotes the distance from the p$^{+}$ edge to the LGAD implant edge. Right: oxide-filled trench isolation -- the trench is characterized by its depth (TD) and width (TW), with spacing measured from the p$^{+}$ edge to the trench edge. The metal, titanium layer (yellow region), oxide passivation, p$^{+}$ anode, and LGAD gain layer are indicated.}
    \label{fig:design_crosssection}
\end{figure}

\begin{figure}[t]
    \centering
    \begin{subfigure}[t]{0.49\textwidth}
        \includegraphics[width=1.0\linewidth]{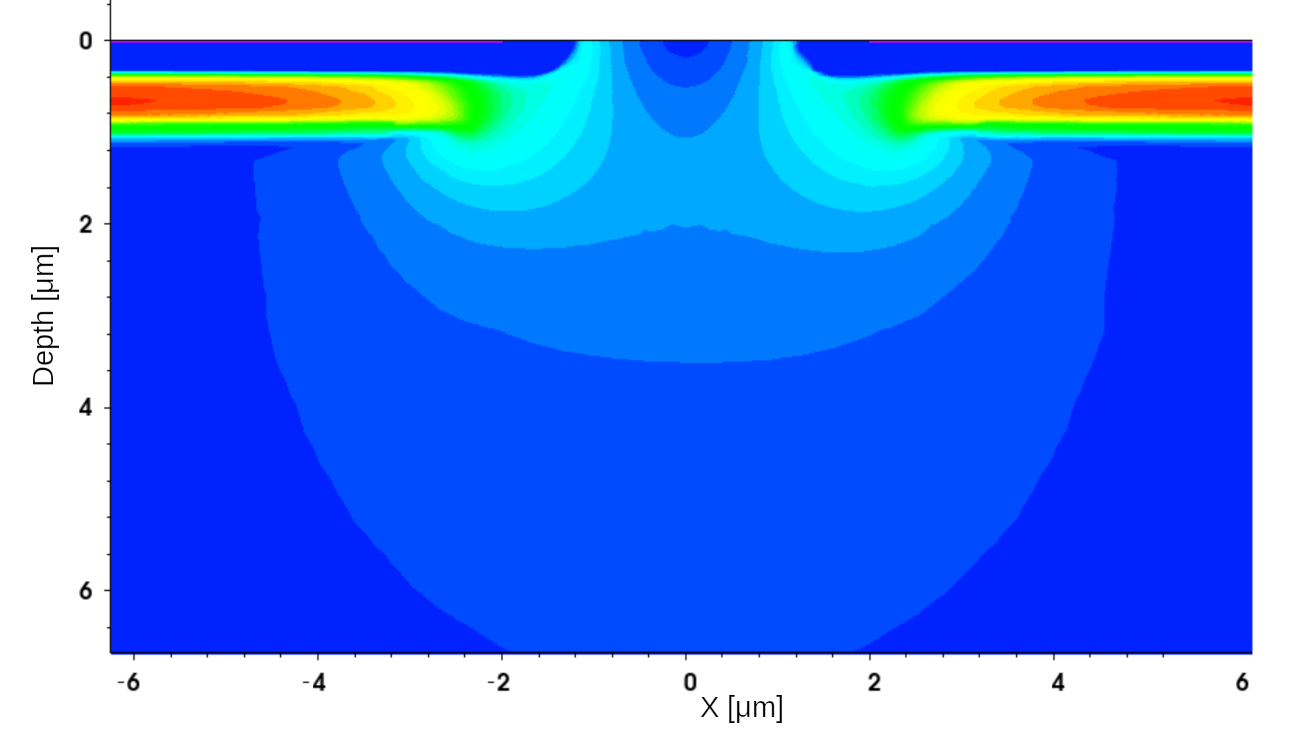}
        \caption{Spacing and gap}
    \end{subfigure}
    \begin{subfigure}[t]{0.49\textwidth}
        \includegraphics[width=1.0\linewidth]{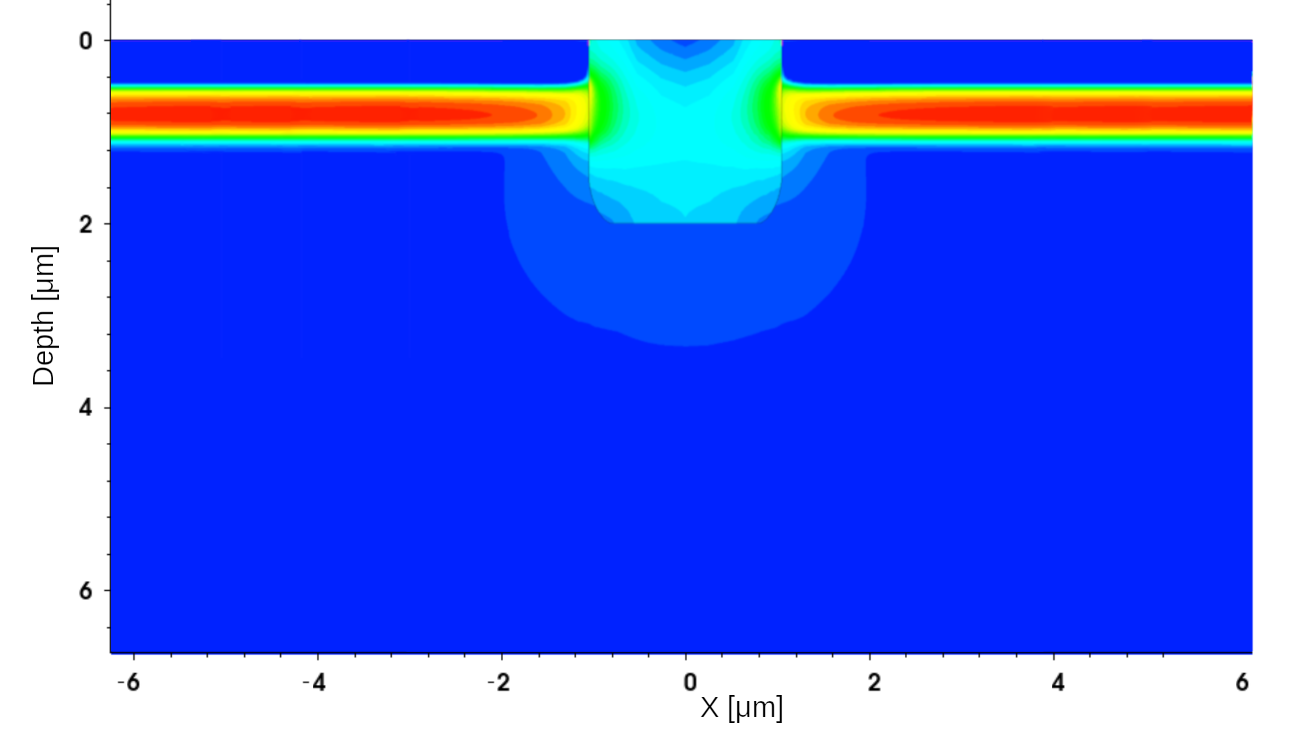}
        \caption{Trench filled with oxide}
    \end{subfigure}
    \caption{Example of LGAD electric field maps from TCAD for separation and trench segmentation solutions. Electric field intensity in the gain region reaches \SI{3e6}{\volt\per\centi\meter}.}
    \label{fig:TCAD_fields}
\end{figure}

\subsection{Segmentation strategies}

To investigate inter-channel isolation efficiency, several separation methods were implemented, ranging from purely geometric separation of the active implants to oxide-filled trenches.

Roughly half of the devices omit the trench entirely, relying on geometric separation alone. Schematic cross-sections of the two approaches are shown in figure~\ref{fig:design_crosssection}. The spacing parameter refers to the lateral distance from the p$^{+}$ implant edge to the trench edge (in trench-segmented variants) or to the geometric mid-plane between adjacent active regions (in trench-less variants). The gap parameter denotes the distance from the p$^{+}$ edge to the LGAD implant edge. In the trench design, the trench depth (TD) and trench width (TW) were additional parameters, however, for simplicity only one was later selected for the fabrication ($TD \geq \SI{0.6}{\micro\meter}$ and $TW = \SI{0.5}{\micro\meter}$). The subset of geometries selected for fabrication, guided by TCAD simulations (example of electric field profile is shown in figure~\ref{fig:TCAD_fields}), is summarized in table~\ref{tab:lgad-split}.

\begin{figure}[tp]
    \centering
    \begin{subfigure}[t]{0.49\textwidth}
        \includegraphics[width=1.0\textwidth]{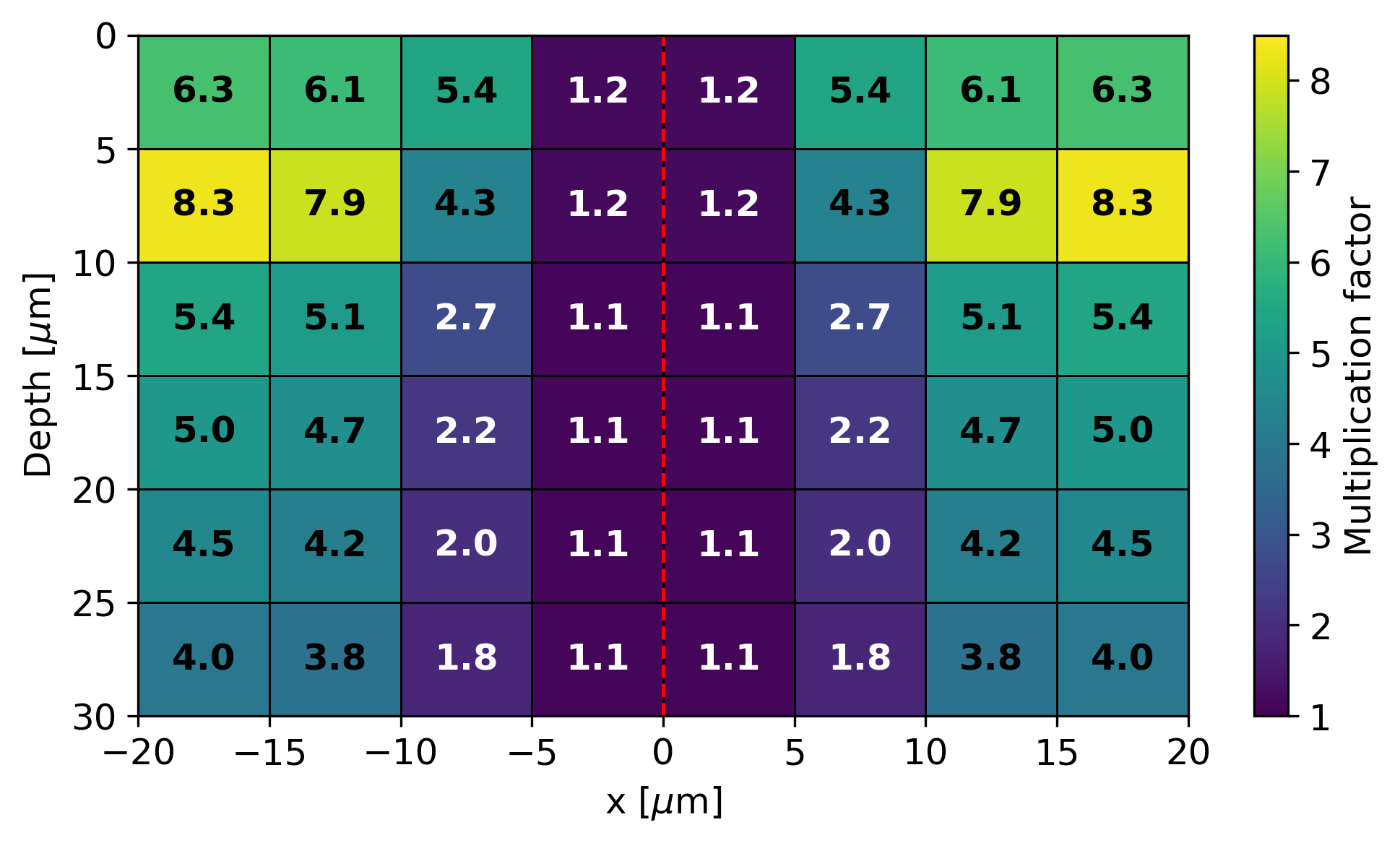}
        \caption{Spacing \SI{0.5}{\micro\meter}, gap \SI{0.5}{\micro\meter}}
    \end{subfigure}
    \begin{subfigure}[t]{0.49\textwidth}
        \includegraphics[width=1.\textwidth]{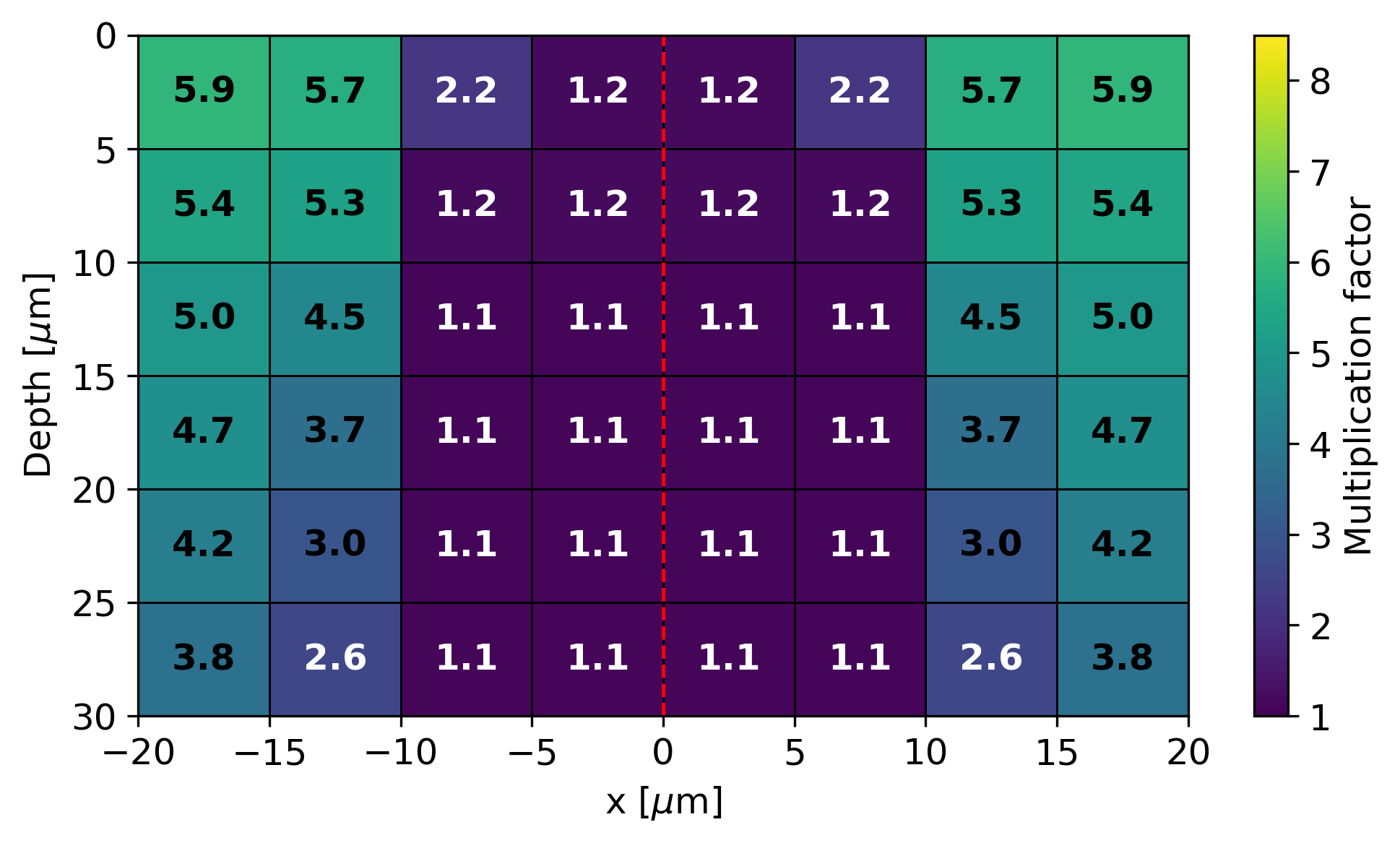}
        \caption{Spacing \SI{0.5}{\micro\meter}, gap \SI{1.5}{\micro\meter}}
    \end{subfigure}
    \\
    \begin{subfigure}[t]{0.49\textwidth}
        \includegraphics[width=1.0\textwidth]{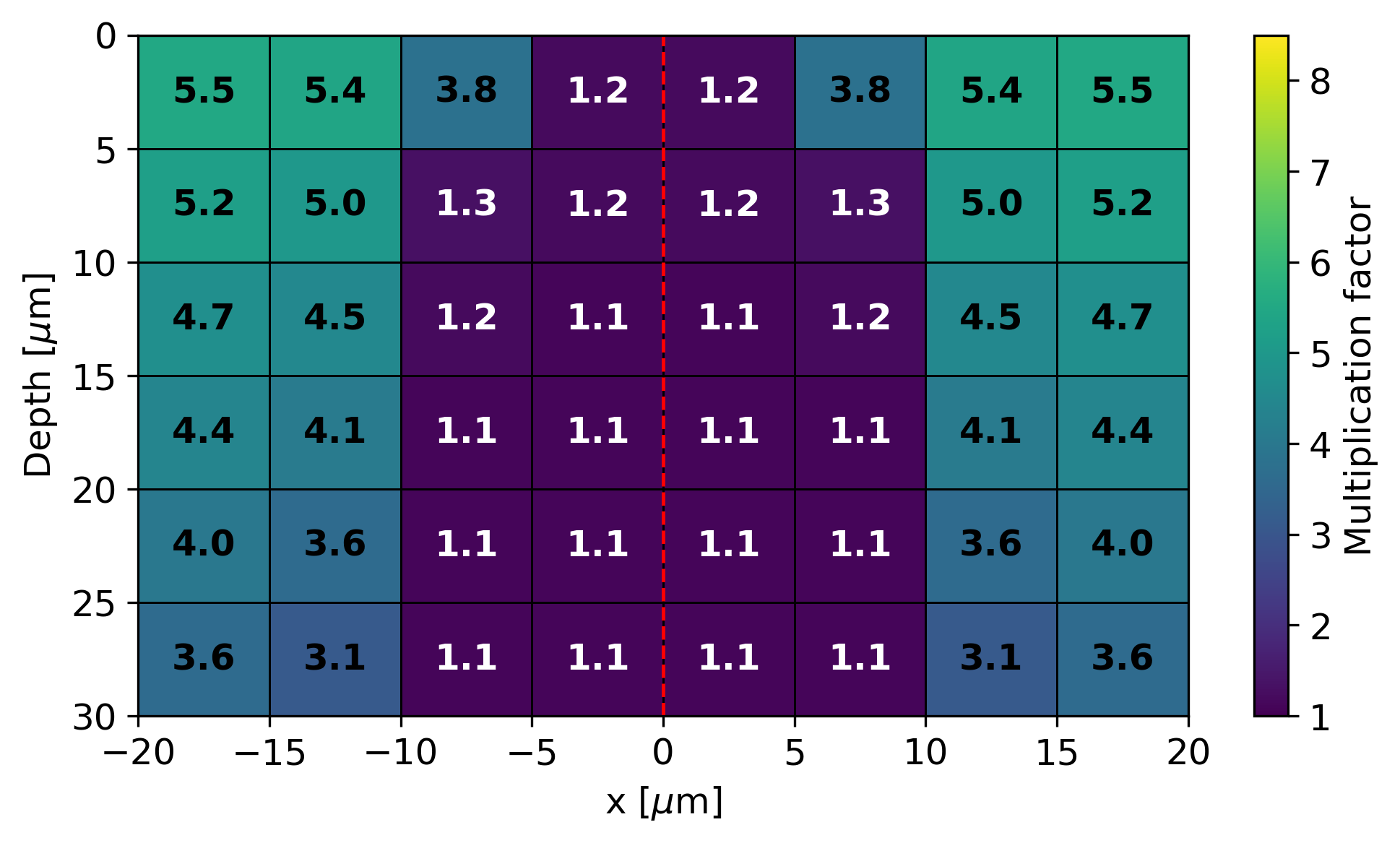}
        \caption{Spacing \SI{1.0}{\micro\meter}, gap \SI{0.5}{\micro\meter}}
    \end{subfigure}
    \begin{subfigure}[t]{0.49\textwidth}
        \includegraphics[width=1.0\textwidth]{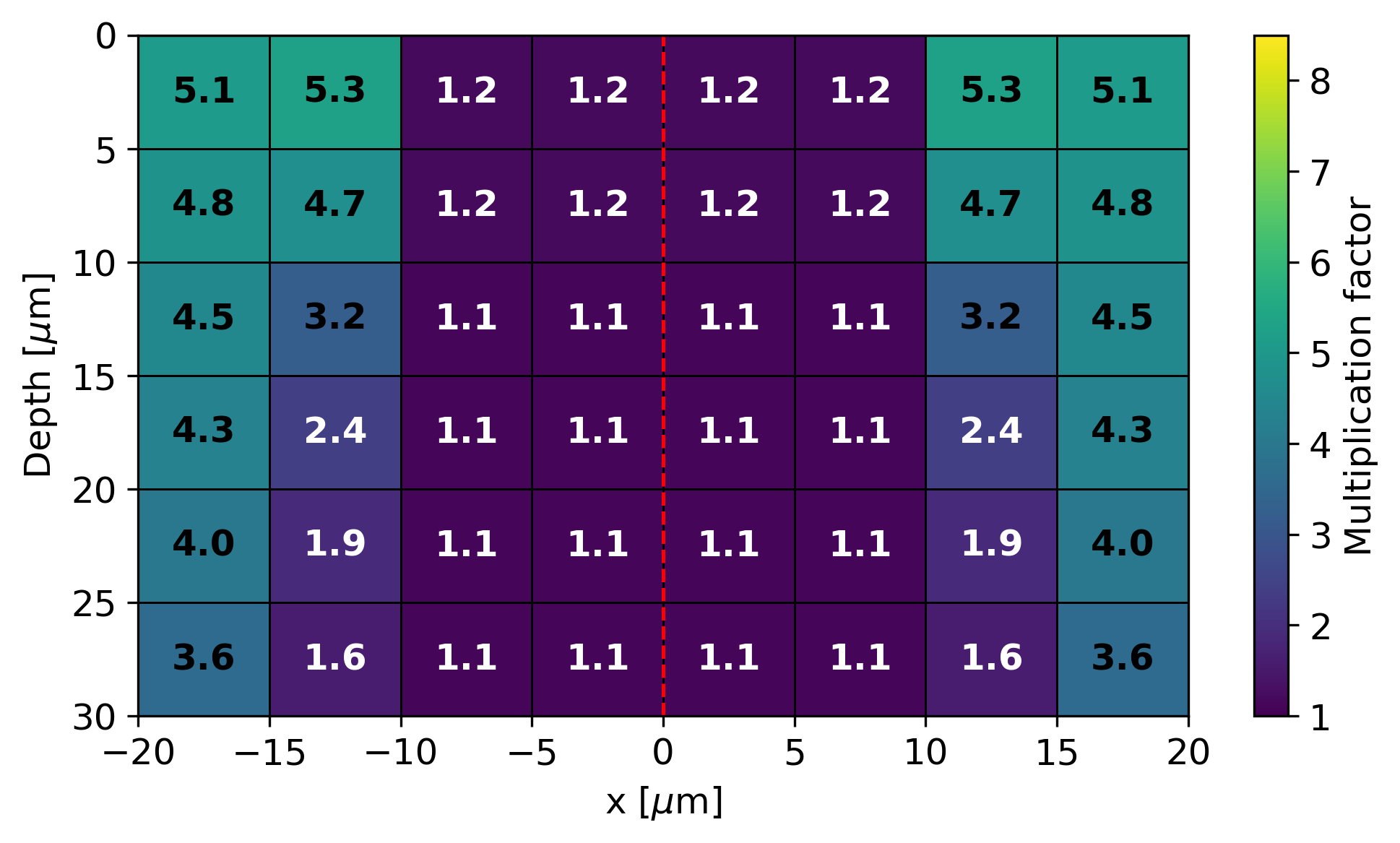}
        \caption{Spacing \SI{1.0}{\micro\meter}, gap \SI{1.5}{\micro\meter}}
        \label{fig:TCAD_gain_separation_measured}
    \end{subfigure}
    \caption{TCAD-simulated multiplication factor for geometrically separated, trench-less SiC LGAD channels at a bias voltage of 300~V for four combinations of spacing $S$ and gap $G$. The multiplication factor is defined as the ratio between the collected and generated charge for a point-like energy deposition corresponding to approximately \SI{80}{\femto\coulomb}. The injection point is scanned on an $(x,y)$ grid through the sensor depth, emulating a TPA-TCT-like measurement. The dashed red line marks the geometrical mid-plane between two adjacent channels at $x=\SI{0}{\micro\meter}$. The multiplication factor reaches values of about 5--8 within the active implant region and falls to approximately unity in the inter-channel region. A larger gap increases the lateral extent of the low-gain region, whereas the spacing parameter changes the position of the active-region edge with respect to the channel boundary. The results highlight the compromise between inter-channel separation and effective gain-layer fill factor in trench-less SiC LGAD segmentation.}
    \label{fig:TCAD_gain_separation}
\end{figure}

\paragraph{TCAD simulations}
The impact of these parameters on the internal gain was studied with TCAD simulations at 300~V bias.
The simulations were done both for laser-like induced signals as well as interaction of minimum-ionizing particles (MIP).

Figure~\ref{fig:TCAD_gain_separation} shows the multiplication factor obtained from TCAD simulations of trench-less, geometrically separated SiC LGAD channels.
The multiplication factor, defined as the ratio of collected to generated charge for a point-like deposition, reaches values of about 5--8 inside the active implant region and decreases to approximately unity in the inter-channel region.
This indicates that the geometrical separation effectively suppresses avalanche multiplication around the channel boundary.

The extent of the low-gain region is governed mainly by the gap parameter: larger gaps increase the lateral size of the inactive region and therefore reduce the effective gain-layer fill factor.
The spacing parameter changes the position of the active-region edge relative to the geometrical mid-plane between channels.
The trench-less geometry therefore provides good suppression of gain between channels, but introduces a sizeable no-gain region, motivating the comparison with trench-isolated SiC LGAD designs.

\begin{figure}[tp!]
    \centering
    \includegraphics[width=1\linewidth]{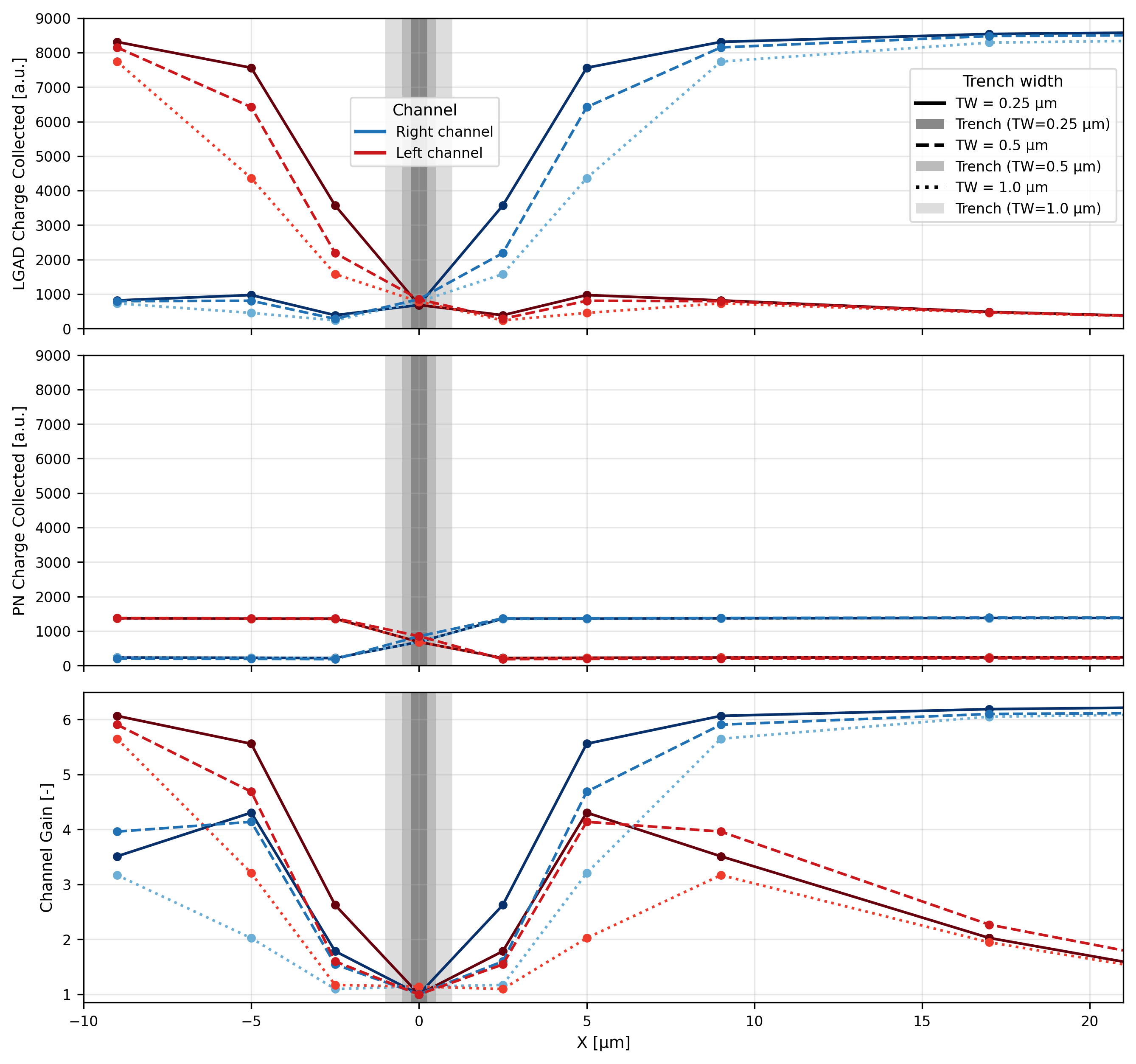}
    \caption{TCAD-simulated response of two adjacent trench-segmented SiC LGAD channels to a perpendicularly incident MIP as a function of the lateral impact position $X$, at a bias voltage of 300~V. The top and middle panels show the charge collected by the LGAD and reference PN structures, respectively, while the bottom panel shows the corresponding LGAD-to-PN charge ratio, interpreted as the effective channel gain. The channel spacing and gap are set to zero and the TD is \SI{0.6}{\micro\meter}; only the oxide-filled trench width is varied between \SI{0.25}{\micro\meter} and \SI{1.00}{\micro\meter}. The gray bands indicate the trench positions for the different TW values. Red curves correspond to the left channel and blue curves to the right channel; line style and color shade distinguish the trench width. Away from the trench, the channel located below the MIP reaches an effective gain of approximately 6. Near the trench, the gain decreases to approximately unity and the collected signals cross over between the two channels, demonstrating charge sharing across the isolation region. Increasing the trench width broadens the low-gain region around the channel boundary and reduces the overlap of the amplified response between neighboring channels, thereby improving electrical isolation at the cost of a larger inactive region.}
    \label{fig:TCAD_gain_trench}
\end{figure}

Figure~\ref{fig:TCAD_gain_trench} shows the TCAD-simulated charge sharing between two adjacent trench-segmented SiC channels for a perpendicularly incident MIP. The comparison between the LGAD and PN structures separates the
effect of geometrical charge sharing from the additional amplification provided by the gain layer. In the PN reference structures, the collected charge changes smoothly between the two channels and shows only a weak dependence on the trench width. In contrast, the LGAD response is much more sensitive to the trench geometry because the local electric-field configuration near the oxide-filled trench affects the multiplication region.

For impact positions well inside a channel, the SiC LGAD reaches an effective gain of about 6. Around the trench, the gain is suppressed to approximately unity, indicating that charge generated close to the isolation structure is
collected with little or no avalanche multiplication. The transition region extends over several micrometers from the trench, with the exact shape depending on the trench width. Wider trenches reduce the overlap of the amplified signals from neighboring channels and therefore improve inter-channel isolation, but they also broaden the no-gain region at the channel boundary. This illustrates the main design trade-off for trench-segmented SiC LGADs: improved segmentation and reduced charge sharing must be balanced against the loss of active gain area.

\section{Initial results}
\label{sec:results}

\subsection{Electrical characterization}
\label{sec:characterization}
Reverse bias measurements (IV and CV) were performed on a representative set of strip-segmented devices from Lot 4 to evaluate the fabrication yield and baseline electrical characteristics. The measurements were carried out using a Karl Suss PA200 probe station, a Keithley 237 Source Measure Unit (SMU), and an HP/Agilent 4284A Precision LCR meter. During the measurements, the strip sensors were wire-bonded to CERN C1 PCBs. The PCB was placed on the probe station chuck, with the HI HV terminal connected to the backside of the sensor. The LO HV terminal was connected to the strip electrodes via direct contact between the probe needle and the PCB pad to which the strips were wire-bonded.

\begin{figure} [tp!]
    \centering
    \begin{subfigure}{0.49\textwidth}
        \includegraphics[width=1.\linewidth]{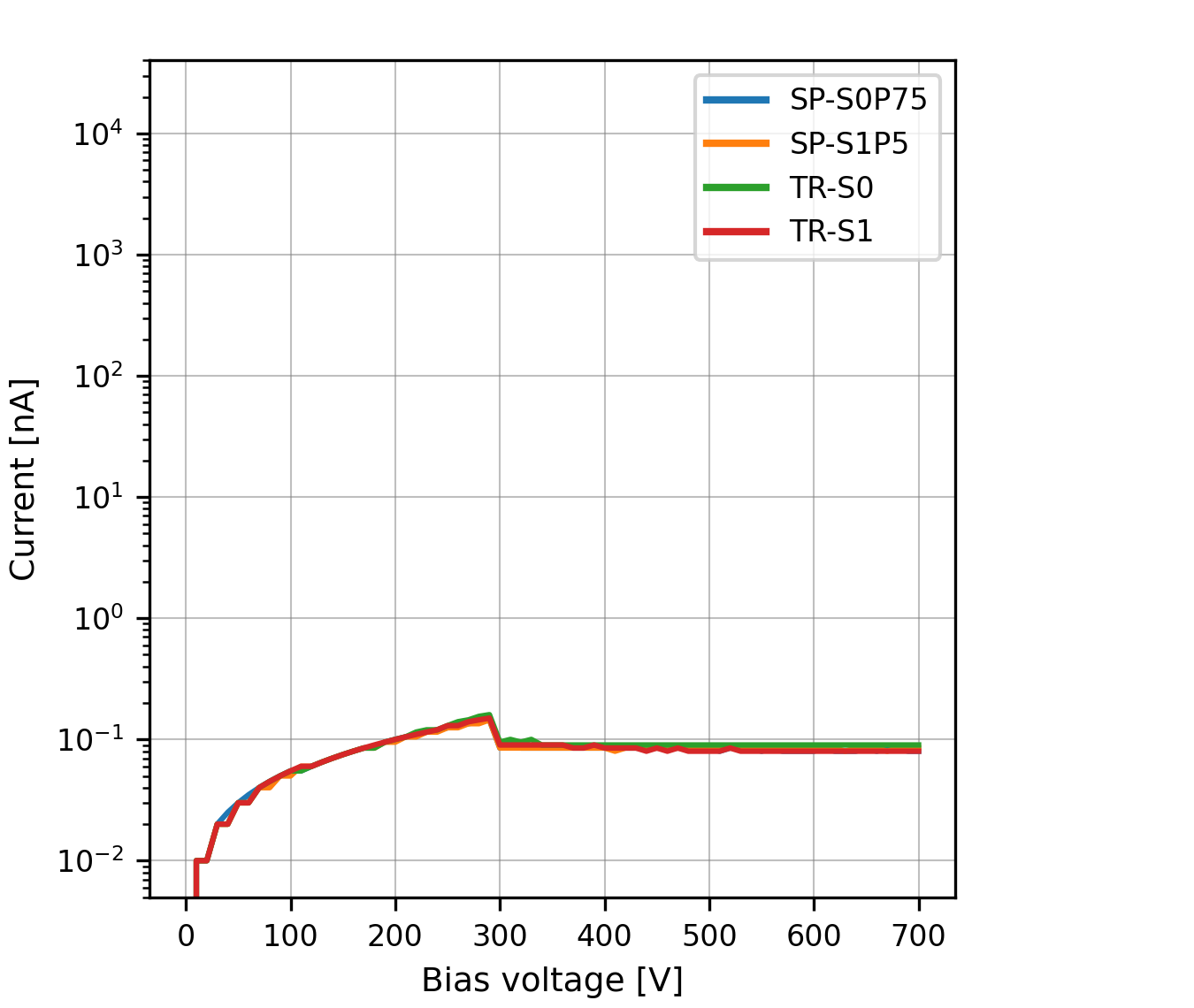}
        \caption{IV}
    \end{subfigure}
    \begin{subfigure}{0.49\textwidth}
        \includegraphics[width=1.\linewidth]{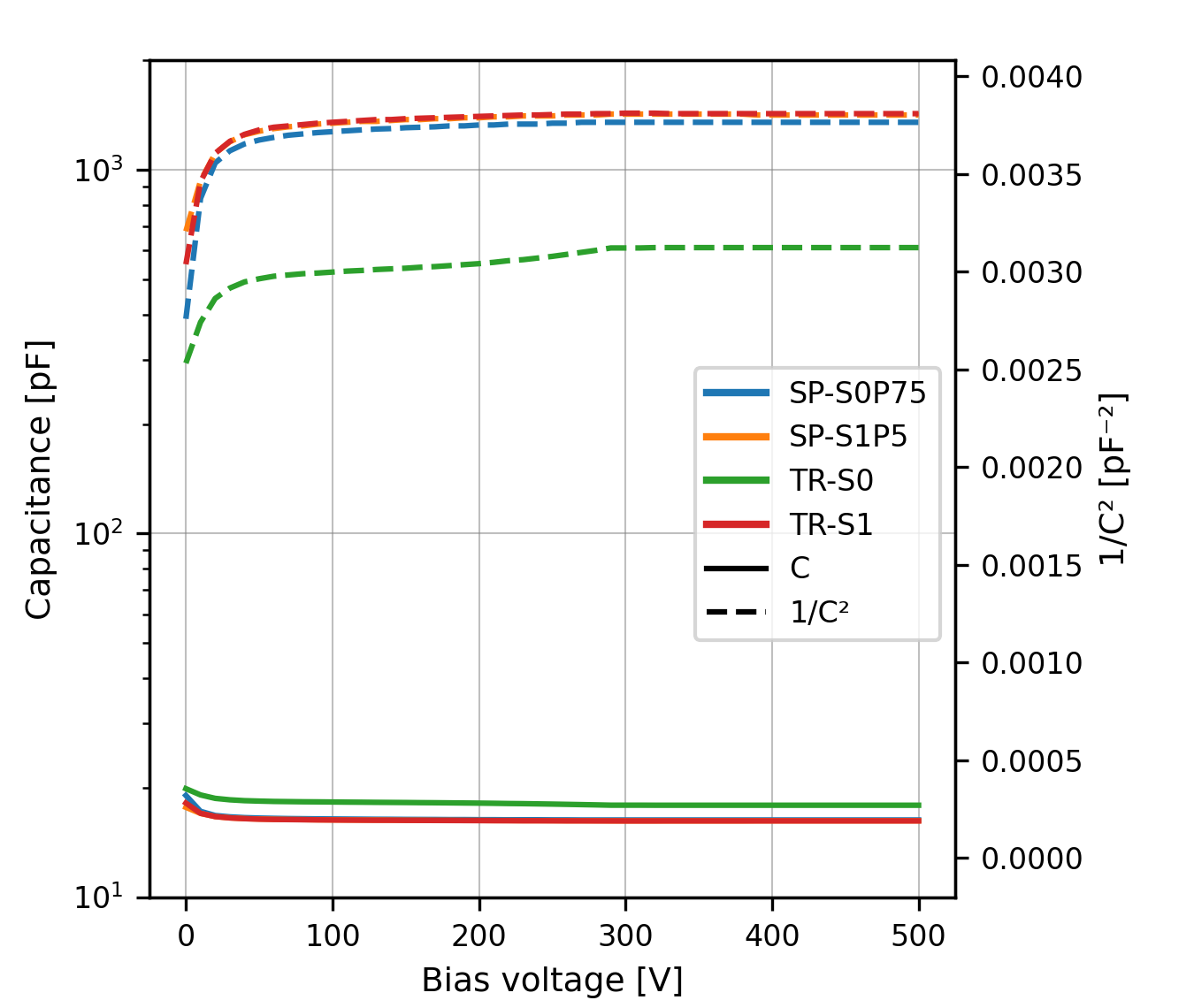}
        \caption{CV}
    \end{subfigure}
    \caption{Reverse bias characteristics of PN reference strip devices without a gain layer, for both geometric spacing (SP) and trench (TR) isolation. Legend labels denote isolation type and spacing parameter in \si{\micro\meter}. All devices show stable leakage of approximately 0.1~nA up to 700~V with no breakdown observed. The discontinuity in TR-S0 near 300~V is a suspected instrumentation artifact from SMU range switching.}
    \label{fig:IV_CV_PN_BOTH}
\end{figure}

The PN reference devices (figure~\ref{fig:IV_CV_PN_BOTH}) provide the electrical baseline against which the LGAD characteristics are compared. All PN diodes exhibit stable leakage currents of approximately 0.1~nA up to 700~V of reverse bias, with no indication of breakdown across both isolation strategies. The CV characteristics show a smooth depletion profile without the two-step structure observed in the LGAD devices. A discontinuity in the IV curves near 300~V is a suspected instrumentation artifact from the SMU switching its measurement range during the sweep. 

\begin{figure} [tp!]
    \centering
    \begin{subfigure}{0.49\textwidth}
        \includegraphics[width=1.\linewidth]{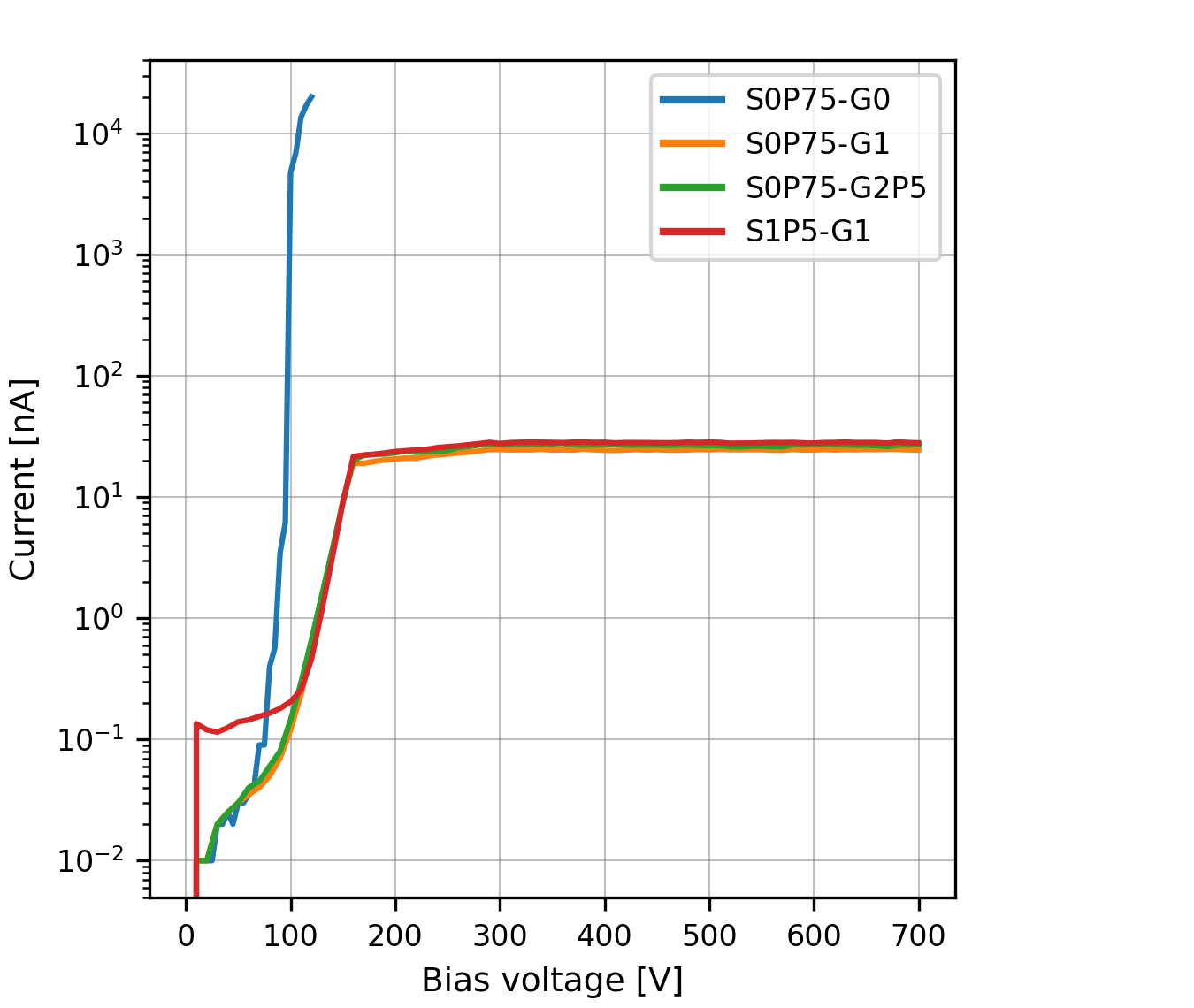}
        \caption{IV}
    \end{subfigure}
    \begin{subfigure}{0.49\textwidth}
        \includegraphics[width=1.\linewidth]{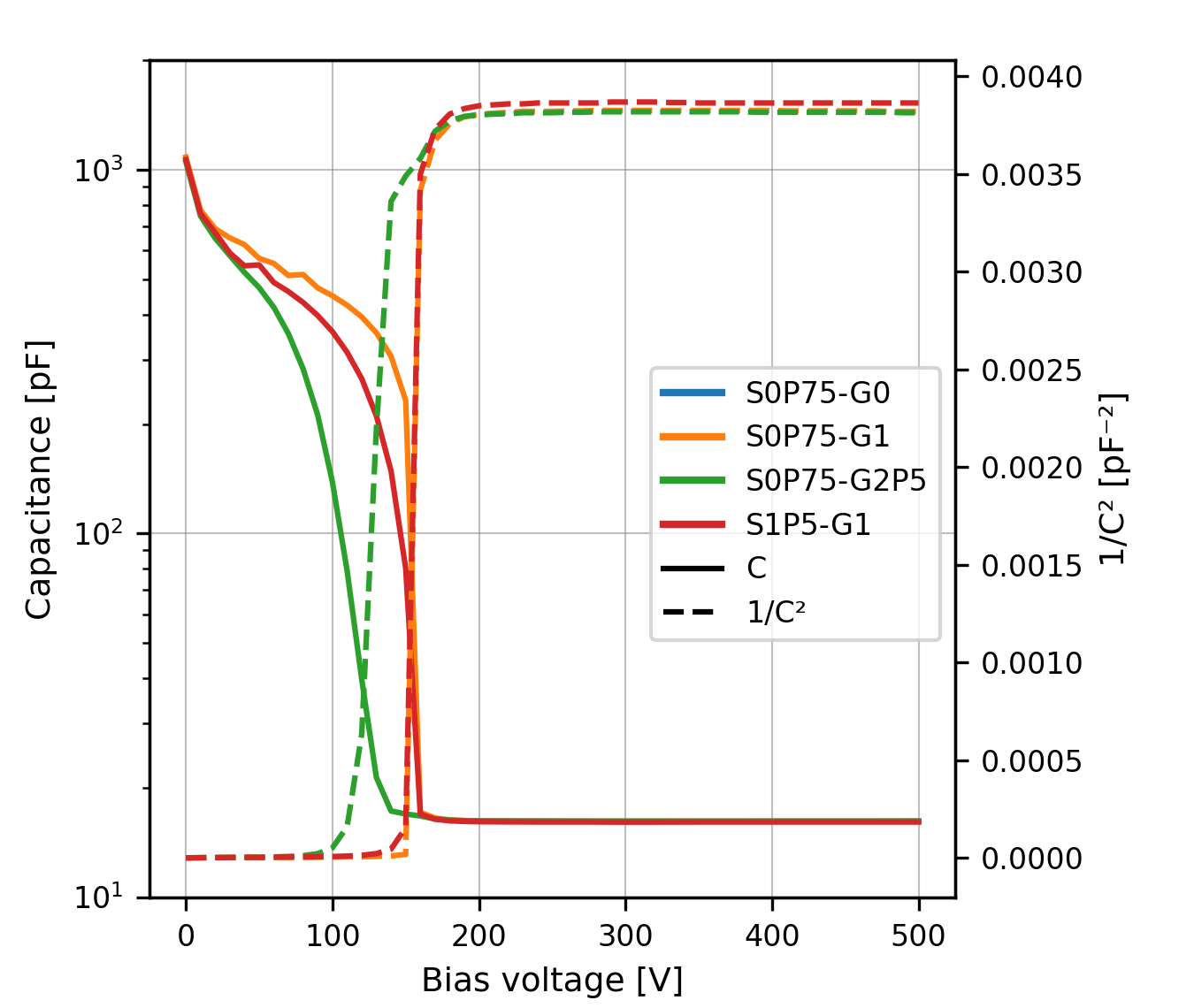}
        \caption{CV}
    \end{subfigure}
    \caption{Reverse bias characteristics of geometrically separated LGAD strip devices. Legend labels denote isolation type and spacing parameter in \si{\micro\meter}. S0P75-G0 exhibits early breakdown near 100~V; the remaining variants show stable leakage of approximately 30~nA up to at least 700~V. The $1/C^2$ knee near 100--200~V marks the depletion of the gain layer. The same gap-dependent breakdown pattern is observed in the trench-isolated variants (figure~\ref{fig:IV_CV_LGAD_TR}).}
    \label{fig:IV_CV_LGAD_SP}
\end{figure}

\begin{figure} [tp!]
    \centering
    \begin{subfigure}{0.49\textwidth}
        \includegraphics[width=1.\linewidth]{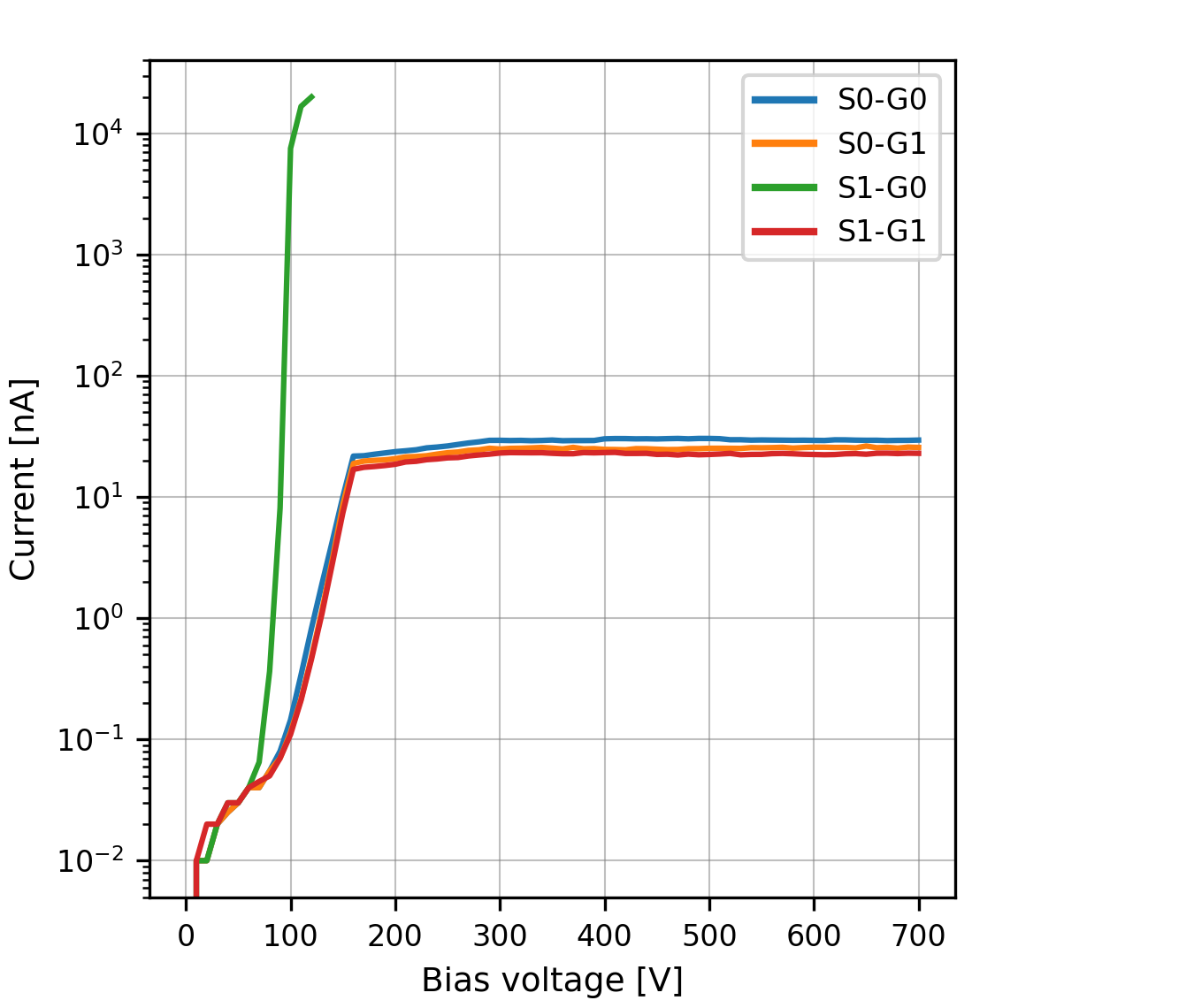}
        \caption{IV}
    \end{subfigure}
    \begin{subfigure}{0.49\textwidth}
        \includegraphics[width=1.\linewidth]{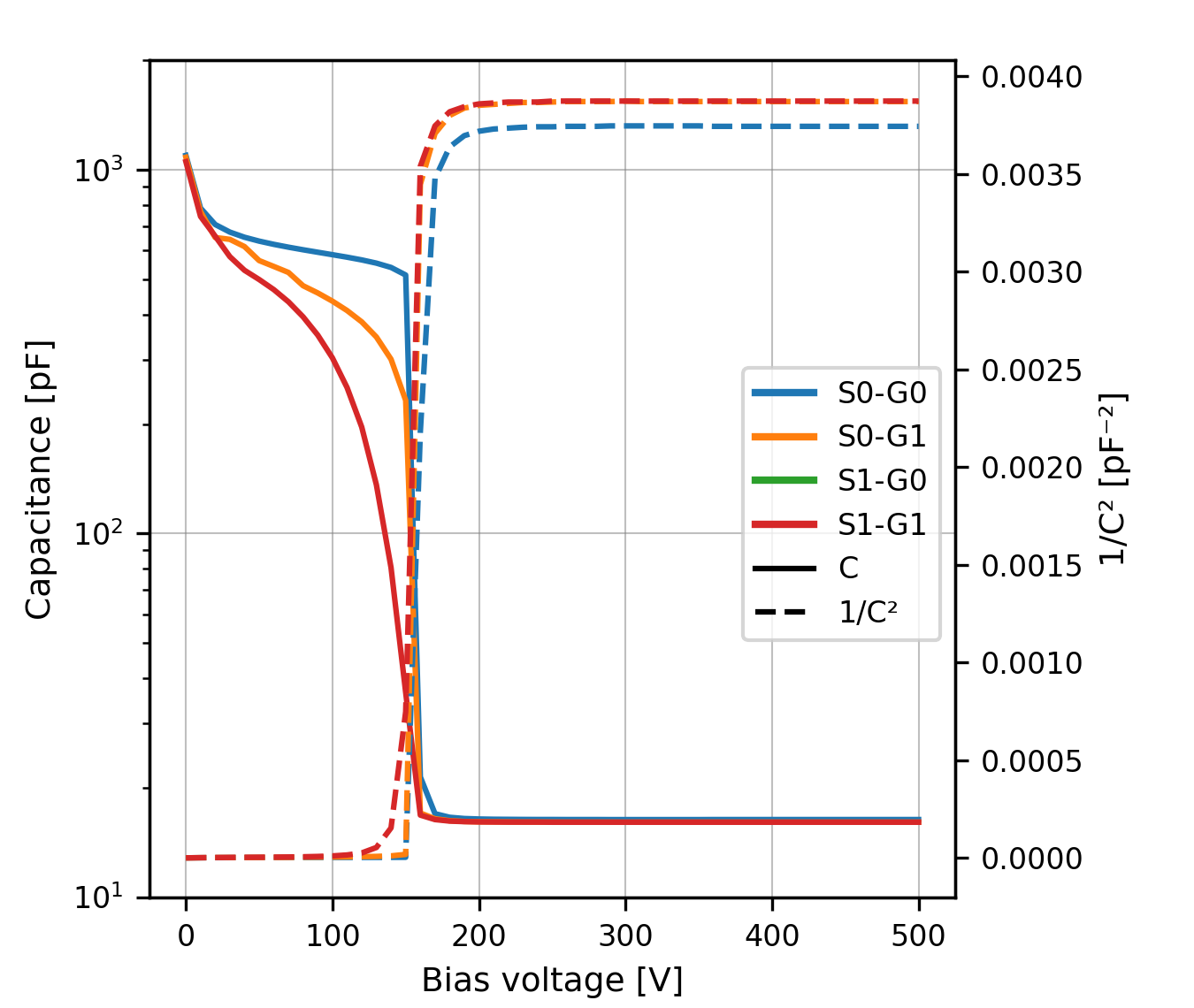}
        \caption{CV}
    \end{subfigure}
    \caption{IV (left) and CV (right) characteristics of trench-isolated LGAD strip devices. Legend labels denote isolation type and spacing parameter in \si{\micro\meter}. Device with gap \SI{0}{\micro\meter} and separation \SI{1}{\micro\meter} (S1-G0) exhibit early breakdown below 100~V, while other devices remain stable up to at least 700~V -- consistent with the gap-dependent pattern observed in the spacing-isolated variants (figure~\ref{fig:IV_CV_LGAD_SP}).}
    \label{fig:IV_CV_LGAD_TR}
\end{figure}

Figures~\ref{fig:IV_CV_LGAD_SP} and~\ref{fig:IV_CV_LGAD_TR} show the IV and CV characteristics of LGAD devices with geometric spacing and trench isolation respectively. The LGAD leakage currents are elevated to the order of 30~nA -- approximately two orders of magnitude above the PN reference -- reflecting the contributions of gain-layer multiplication of thermally generated carriers. The CV curves exhibit a characteristic two-step profile: an initial capacitance decrease associated with depletion of the gain layer, followed by a sharper drop corresponding to depletion spreading into the lightly doped epitaxial bulk. The transition is also visible in the $1/C^2$ curves, where the knee identifies the gain-layer depletion voltage.

The most notable observation across both isolation strategies is the correlation between the gap parameter and the breakdown voltage. Devices with gap \SI{0}{\micro\meter} and spacing~$>0$, exhibit early breakdown below approximately 100~V in both the spacing-isolated (figure~\ref{fig:IV_CV_LGAD_SP}, S0P75-G0) and trench-isolated (figure~\ref{fig:IV_CV_LGAD_TR}, S1-G0) variants. In contrast, devices with gap~$\geq$\SI{1}{\micro\meter} remain stable up to at least 700~V. The consistency of this behavior across both isolation approaches suggests that retracting the gain implant from the channel boundary is necessary to suppress electric-field crowding at the inter-channel interface and avoid premature avalanche breakdown in segmented 4H-SiC LGADs.

In the trench-isolated variants, devices with separation~$=0$ exhibit significant charge sharing between neighboring channels. This behavior is attributed to the oxide-filled trenches being insufficiently deep to fully isolate the multiplication regions electrically. Since no dedicated implantation mask was implemented above the trench region, the gain implant remains continuous underneath the $\geq$~\SI{0.6}{\micro\meter}-deep trench, allowing electrical coupling between adjacent strips. As a result, the trenches provide only partial electrical isolation despite the geometric separation.


\begin{figure}[tp!]
    \centering
    \includegraphics[width=1\linewidth]{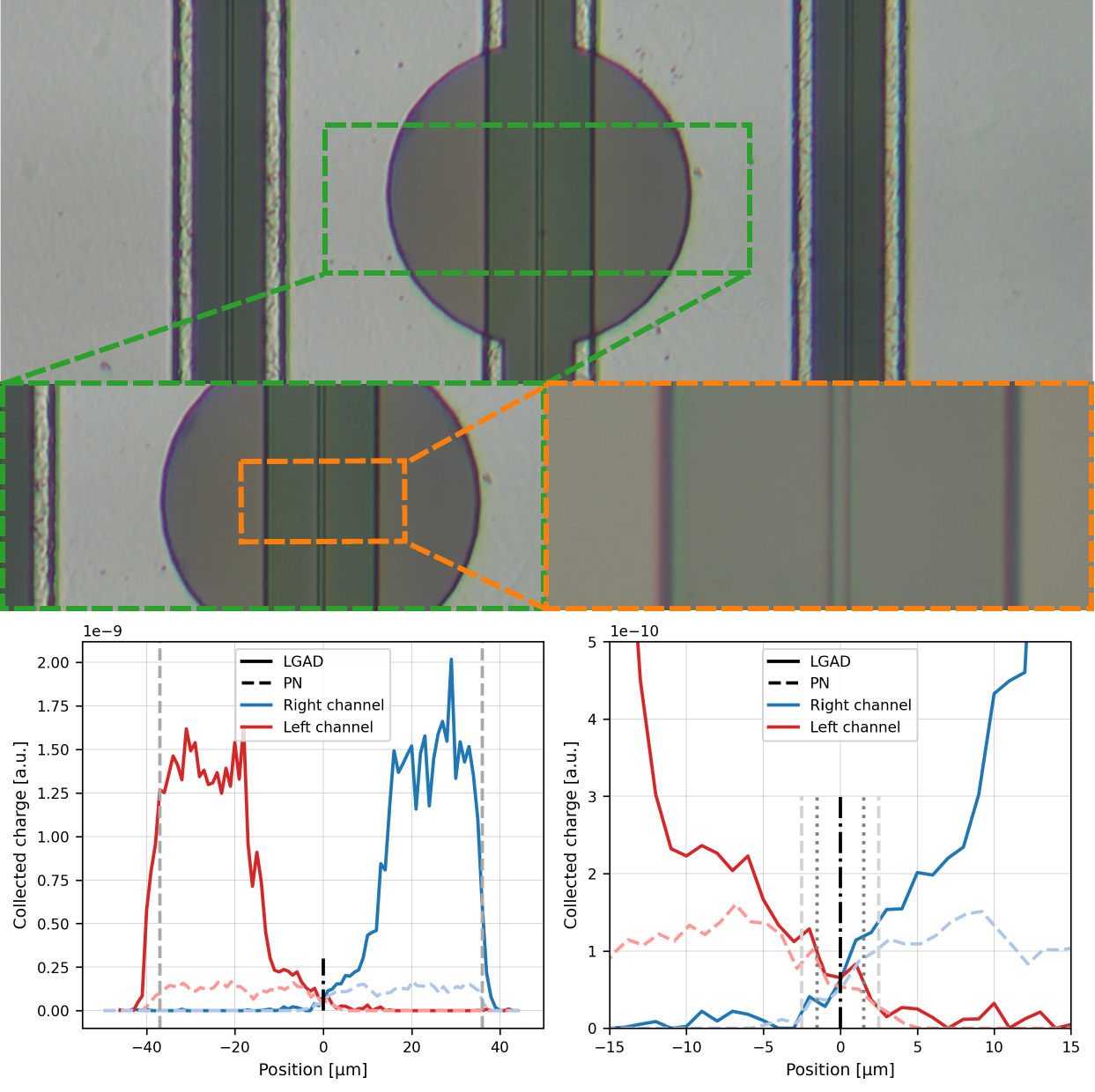}
    \caption{Collected charge as a function of lateral position across two adjacent PN and LGAD strips, measured by TPA-TCT at ELI ERIC at 300~V. The top of the figure shows a micrograph of the circular opening in the strip metallization; magnified insets indicate the regions over which the charge-collection profiles shown below were acquired. Signals from the left channel are shown in red hues and those from the right channel in blue. Left: full scan across both strips for the LGAD and PN devices; the dashed vertical line marks the approximate position of the metal edge, and the dash-dotted line indicates the midplane between the strips. Right: close-up of the inter-strip region; the vertical lines mark the geometric boundaries of the strip separation -- the spacing (dotted) and the gap (dashed). Indicative gain is around 6-10.}
    \label{fig:TCT_LGAD_PN}
\end{figure}

\subsection{TPA-TCT measurements}
\label{sec:TCT}
Two-photon absorption transient current technique (TPA-TCT) measurements were carried out at the ELI ERIC laser facility to probe the charge collection and inter-channel isolation of segmented strip devices.
For the initial tests, the design with the safest spacing \SI{1.5}{\micro\meter} and gap \SI{1.0}{\micro\meter} has been used on the \SI{60}{\micro\meter} metal strip variant with a hole opening.
The TPA-TCT method uses a tightly focused femtosecond laser to generate charge carriers at a well-defined depth (in our case around \SI{20}{\micro\meter}) within the sensor bulk, enabling spatially resolved charge collection mapping across the device.
The signal yield reported in the following is \emph{uncalibrated}: it is expressed in arbitrary units corresponding to the integrated current response of the readout chain consisting of \SI{40}{\decibel}, \SI{2}{\giga\hertz} bandwidth CIVIDEC amplifier and a \SI{2}{\giga\hertz} oscilloscope. The interpretation assumes that the carrier generation is dominated by two-photon absorption at 400~nm laser wavelength.
Only minimal energy scaling has been applied for comparison and neither the single-photon absorption (SPA) nor the conversion to absolute charge or to deposited energy has been applied.
Therefore the results should be read as relative trends rather than absolute yields.

Figure~\ref{fig:TCT_LGAD_PN} shows the collected charge as a function of the lateral position across two adjacent strips for an LGAD device and a PN device of the same geometry.
The TPA signal integrals recorded by the PN were scaled to the same laser power using quadratic law \cite{wiehe2021tpatct} to match the power used during the LGAD acquisition, enabling a direct shape comparison with the LGAD response.
The two readout channels \emph{(Left and Right)} correspond to the adjacent strips.
The measured profiles reproduce the behavior predicted by the TCAD simulations presented earlier -- geometry wise (see figure~\ref{fig:TCAD_gain_separation_measured}), both in the position of the inter-strip transition and in the relative amplitudes of the LGAD and PN responses, providing experimental confirmation of the simulated device geometry and isolation strategy.

In the LGAD device, clear charge collection is observed within each strip, with a sharp transition in the inter-strip region.
The signal amplitude is notably higher than in the PN device, consistent with the internal gain provided by the multiplication layer; the ratio between the two profiles is compatible with the gain expected from simulation, within the limits imposed by the uncalibrated measurement and the assumption of pure two-photon generation.
The detail view (figure~\ref{fig:TCT_LGAD_PN}, right) reveals the charge-sharing behavior in the gap between strips: the crossover between the two channels occurs within a few micrometers of the geometric mid-plane, demonstrating effective electrical isolation between adjacent segments and in good agreement with the simulated fill factor (ratio of gain and no-gain area).
The dashed and dotted vertical lines mark the geometric boundaries of the separation structure -- spacing, gap, and metal edge -- allowing the measured profiles to be correlated directly with the physical layout of the inter-strip region.

\section{Conclusions and outlook}
\label{sec:conclusions}

Segmented 4H-SiC LGAD devices -- strip detectors with \SI{80}{\micro\meter} pitch and pixel arrays with \SIlist{55;110}{\micro\meter} pitches -- have been designed, fabricated, and initially characterized. These represent the first segmented 4H-SiC detectors incorporating an internal gain layer. Multiple inter-channel isolation strategies were explored, including geometric separation and oxide-filled trenches with various spacing and gap parameters. Initial TPA-TCT exploratory measurements confirm clear charge separation between adjacent strips with internal gain, demonstrating that the segmentation concept established in silicon LGADs can be successfully transferred to 4H-SiC.

Several directions for further development are being pursued. Future work will combine more detailed spatially resolved TPA-TCT scans with the ongoing analysis of MIP data collected at the CERN SPS test-beam facility. Dedicated TCT campaigns are also planned to systematically compare the trench-isolated and geometrically separated designs. The gain layer implantation profiles observed in Lot~4 indicate that the implant parameters require further optimization to match the design targets; adjusted strategies for subsequent wafer production are being developed. Bump-bonding of the \SI{55}{\micro\meter} pitch pixel devices to the Timepix4 readout chip will be pursued as a path towards a fully integrated 4H-SiC pixel detector module.

\acknowledgments
The initial TCT experiments were performed at the Extreme Light Infrastructure ERIC (ELI ERIC) Facility. The authors acknowledge the support of M.~Rebarz and other ELI staff during the experimental campaign. The authors also thank S.~Onder, D.~Radmanovac, and S.~Gundacker from the Marietta Blau Institute for Particle Physics (MBI), Vienna, for their substantial help in acquiring the initial TCT measurements used in this work.

This work was supported by the Technological Agency of the Czech Republic -- Project TK05020011 and by project LM2023040 CERN-CZ .
Researcher Peter Švihra conducts his research under the Marie Skłodowska-Curie Actions – COFUND project, which is co-funded by the European Union (Physics for Future – Grant Agreement No. 101081515).
The team from the Institute of Physics of the Czech Academy of Sciences was also supported via the project FORTE - CZ.02.01.01/00/22\_008/0004632.

\section*{Declaration of Generative AI and AI-assisted technologies in the writing process}
During the preparation of this work the author(s) used Claude in order to improve the readability of the paper. After using this tool/service, the author(s) reviewed and edited the content as needed and take(s) full responsibility for the content of the published article.

\bibliographystyle{JHEP}
\bibliography{biblio.bib}

\end{document}